      [1999/12/01 v1.4c Il Nuovo Cimento]
\documentclass{cimento}
\usepackage{epsfig,psfig}
\title{Chacaltaya: towards a solution of the knee ....?}
\author{O.~Saavedra\from{ins:x}
        \atque
L.~Jones\from{ins:evil}}
\instlist{\inst{ins:x} Dipartimento di Fisica Generale Torino, Italy.\\
Istituto Nazionale di Fisica Nucleare, Torino,Italy.
  \inst{ins:evil} Department of Physics, University of Michigan, 
Ann Arbor, USA.}
\PACSes{\PACSit{96.40}{Cosmic Rays}
\PACSit{96.40.De}{Composition, energy spectra and interactions}
\PACSit{96.40.Pq}{Extensive Air Showers}}
\begin{document}

\maketitle

\begin{abstract}
Cosmic rays physics is currently being studied with rather sophisticated
detectors
running in a variety of experimental conditions and atmospheric depths
around the world. In this paper we describe the reasons why cosmic ray physics
experiments
at high altitudes like Chacaltaya are so important for resolving some of the
open problems in
cosmic physics. A discussion on the
future prospects of the high altitude mountain laboratories such as Chacaltaya
for cosmic ray physics
is presented.
\end{abstract}

\section{Introduction}
One of the most important problems in cosmic ray physics, not yet resolved after
nearly 40 years,
is the existence of the "knee in the primary spectrum.
And there are many other important
problems in the field of cosmic rays that should be studied, in particular
experimental
investigations, and at high altitude sites.\\
One may ask why are observations of cosmic rays at high altitude laboratories
are so important? One reason is because it is possible
to observe the air shower cascades produced by lower energy of primary cosmic
rays than at lower altitudes, due to their attenuation in the atmospheric
overburden.
This is particularly relevant in the current studies of gamma ray sources and
gamma bursters.
Also, to study the shower cascades of higher energies at an early stage of
development
where showers present a minimum of fluctuations.
The atmospheric depth of Chacaltaya (540 $g/cm^2$) corresponds to the maximum
development of the
showers in the energy range of 10 - 500 TeV, giving a maximum detection
probability.
Moreover the fluctuations in the development of the showers are much lower
at the observation level of Chacaltaya than at lower altitudes.\\
In this paper a possible experiment using a variety of different
techniques and carried out by a wide international collaboration is also
presented and discussed.
\section{The Chacaltaya Laboratory}
The Chacaltaya Cosmic Ray Research Laboratory, near La Paz, Bolivia, is located
at an elevation
of 5220 meters above sea level corresponding to $540 gr/cm^2$, equivalent to
about 7 nuclear
interaction mean free paths (for TeV energy protons) and
14.1 radiation lengths. Its geographic position is at $16^\circ$ South Latitude.
$291.8^\circ$ East Longitud. and $-4^\circ$
of geomagnetic latitude corresponding to 13.1 GV cutoff rigidity. Its geographic
location
in the Southern Hemisphere is also important in order to observe a part of the
sky where $\gamma$-astronomy
is not as well developed as it is in the Northern Hemisphere. The Galactic
Center, in particular,
can be observed at low zenith angle.\\
The Chacaltaya Laboratory is the highest continuously functioning research
station on the globe, and provides
a unique opportunity for research on cosmic ray phenomena.
At energies above $10^{14} eV$, the flux of primary cosmic rays is so low
that direct observation by balloon -or satellite-borne instruments (with
areas of only a few square meters) is not feasible.  For example, the integral
primary
cosmic ray flux of energies above $10^{16} eV$ is only one particle per ($m^2
\cdot sr \cdot yr$).
Consequently, for the most sensitive indirect studies of cosmic rays with
energies of
and above $10^{15} eV$ (one PeV), it is necessary to deploy extensive detector
systems at
as high an elevation as possible, to reduce the of atmosphere overburden.

\subsection{Hadronic intensities at Chacaltaya}
A experiment using emulsion chambers in combination with the extensive air
shower
technique is being operated at Chacaltaya by the SYS collaboration. The
collaboration is observing
bundles of high energy gammas and hadrons in the air shower cores (with emulsion
chambers)
associated with air showers detected by the EAS array.
The experiment employs 35 plastic scintillators,
32 emulsion chamber units and a hadron calorimeter.
Each emulsion chamber unit, with dimensions of $50~cm\cdot50~cm\cdot15~cm$,
consisting of 15 layers of 1 cm Pb plates
and two sheets of X-ray films.
Below each unit a scintillator detector of the same area as the
emulsion chamber detects the bundle of charged particles produced in the
emulsion chamber material by the hadronic component of the shower through
local nuclear interactions.\\

The gamma "family" is connected to the hadronic component through the
geometrical
position, and the hadronic component is related to the air shower
through the arrival time~\cite{ref:kawa}.
Each unit of the hadron calorimeter provides an output which is a 
measure of the
energy released in the scintillator and is converted to the number of
(minimum-ionizing) charged particles.
The data produced by the air shower array and by the hadron calorimeter are
recorded when at least one unit of the hadron
calorimeter records a particle density $n_b \geq 10^3$ (particles/0.25 $m^2$).
2408 events were selected during 4.6 years of data taking, during which
the emulsion chambers were simultaneously active.
The shower size interval for these data is $N_e =5 \cdot 10^6 - 10^7$\\
The receintly published results~\cite{ref:agui} can be seen in figure 1 where
the differential energy spectrum of hadrons in air showers is compared with
the expectations from simulations. These simulations used
different models to simulate multiple particle production;
the UA5 algorithm (modified for
hadron-nucleus collisions), VENUS~\cite{ref:wern}, QGSJET~\cite{ref:kalm} and
HDPM~\cite{ref:capd}.
Atmospheric diffusion of cosmic rays is described by the code in
ref.~\cite{ref:niih} for the UA5 model,
and by CORSIKA 5.20~\cite{ref:heck} for the rest.
It is apparent that the average number of hadrons for shower size
$N_e =5 \cdot 10^6 - 10^7$
is lower than that predicted by all these simulations.
This tendency is consistent not only with the relationship between
the $\gamma$ families and the accompanied air showers
discussed by Kawasumi et al.~\cite{ref:kawa},
but also with the conclusions of the KASCADE experiment at sea level
~\cite{ref:risse},
where they have also reported a hadronic flux accompanying
air showers lower than the expectations based on several models.
In  figure 2 the measured muon trigger rate as a function of
the hadronic rate is compared with the
predictions from several simulations for multiple particle production.
In a more recent paper by the KASCADE group~\cite{ref:haun}, this result
is confirmed.
\\
These measurements:
$\gamma$ families and hadrons with the SYS experiment at Chacaltaya,
and hadrons at the KASCADE array at sea level, are quite independent, as they
are observing with different detector systems. However they display similar
problems in describing shower development through the atmosphere.
If we take into account these results,
(i.e. lower hadronic intensities than expected in the KASCADE
experiment~\cite{ref:risse} and at Chacaltaya~\cite{ref:agui},
as well as lower intensities of $\gamma$-families~\cite{ref:kawa}),
it would seem that at
primary energies in the range $10^{15}$-$10^{16}$ eV a larger
dissipative mechanism occurs in nuclear interactions than current models
predict.
In fact, if we consider cosmic
phenomenology, we see the ``knee'' problem seen in the energy spectrum of
primary cosmic rays in the same energy range of the emulsion chamber
families (e, $\gamma$).\\

\begin{figure}[hp]
\begin{minipage}{.49\linewidth}
\begin{center}
\mbox{\epsfig{figure=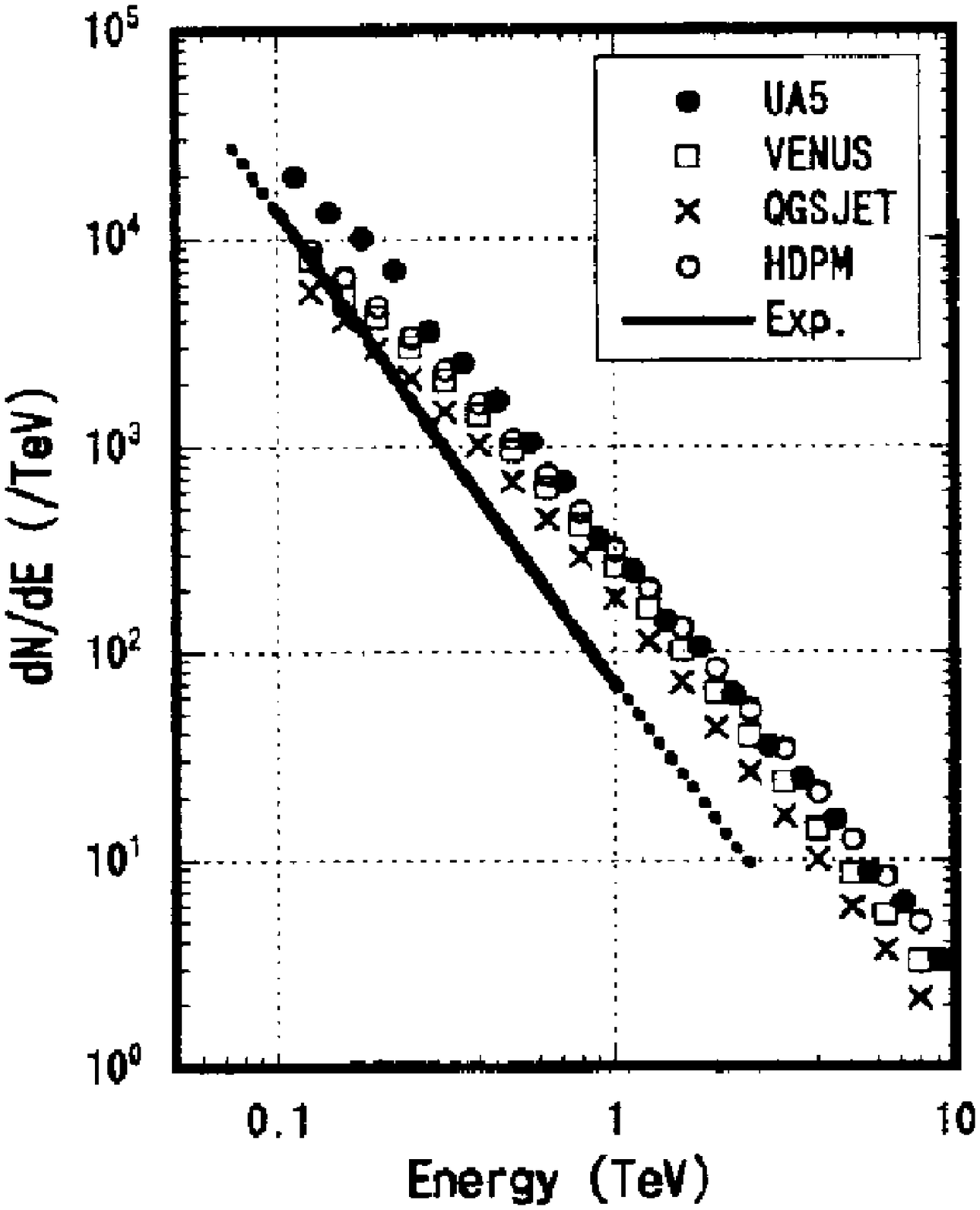,width=6.5cm,height=7.cm}}
\caption{\em {Differential energy spectrum of hadrons in air showers
(solid line) and expectations from MC simulations with various
interaction models (See text). 
The EAS size range is $N_e =5 \cdot 10^6 - 10^7$.}}
\end{center}
\label{noh_e}
\end{minipage}\hfill
\begin{minipage}{.49\linewidth}
\begin{center}
\mbox{\epsfig{figure=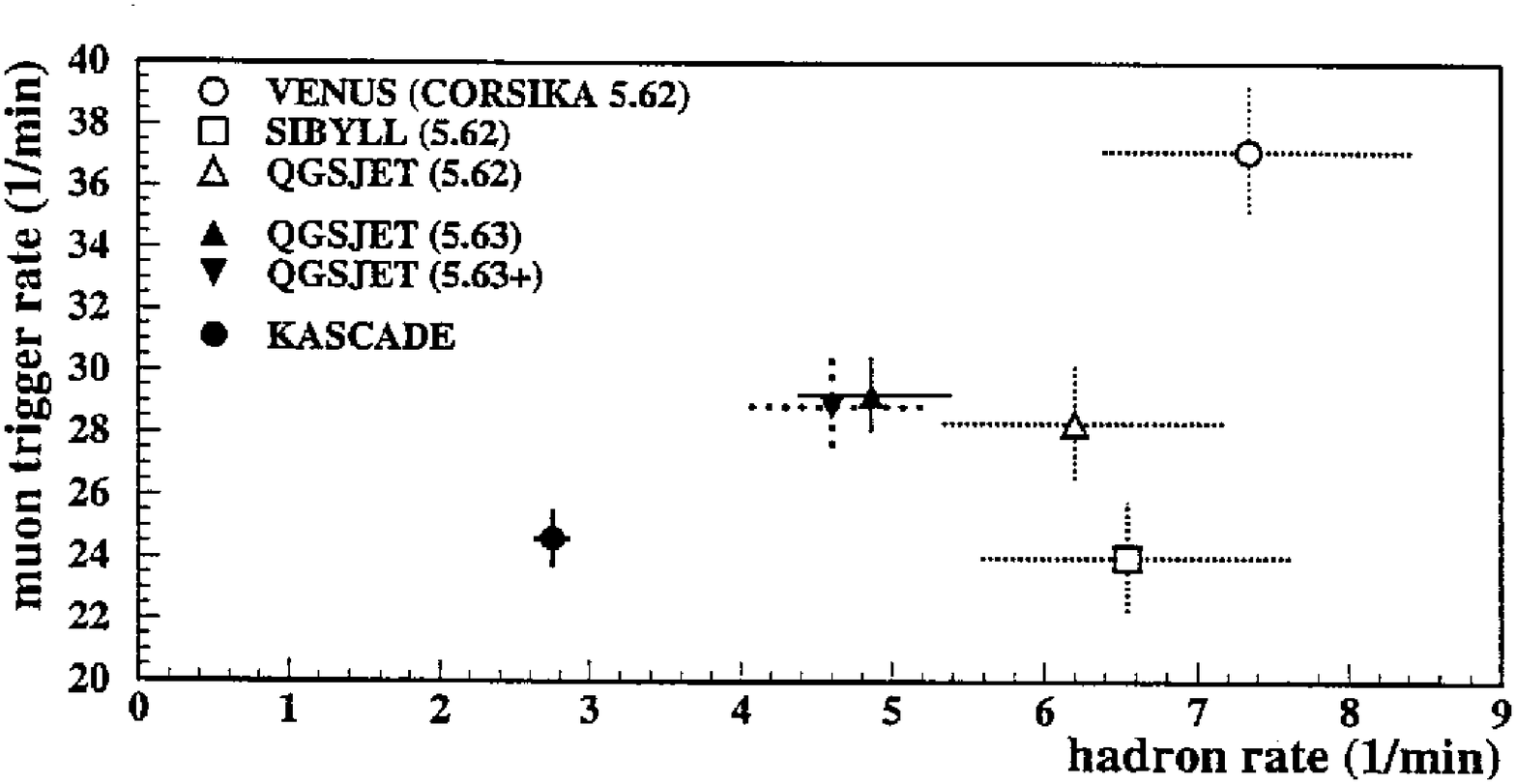,width=6.5cm,height=7.0cm}}
\vspace{-0.3cm}
\caption{\em {Comparison between simulated and measured integral muon trigger
and hadronic
rates at the KASCADE calorimeter. The experimental hadronic rate
is much lower than the expectations by
various simulation codes.}}
\end{center}
\label{mu_had}
\end{minipage}
\end{figure}
\vspace{-0.5cm}

\subsection{Hadronic interaction at colliders and cosmic rays}
The primary cosmic ray spectrum is derived from flux measurements of EAS for
which energies are measured by the total number of particles in the shower.
It is obvious that such measurements at energies
higher than the ``knee'' region can be seriously affected if the
characteristics
of the hadronic interaction change radically.\\

\begin{figure}[htbp]
\begin{center}
\mbox{\psfig{figure=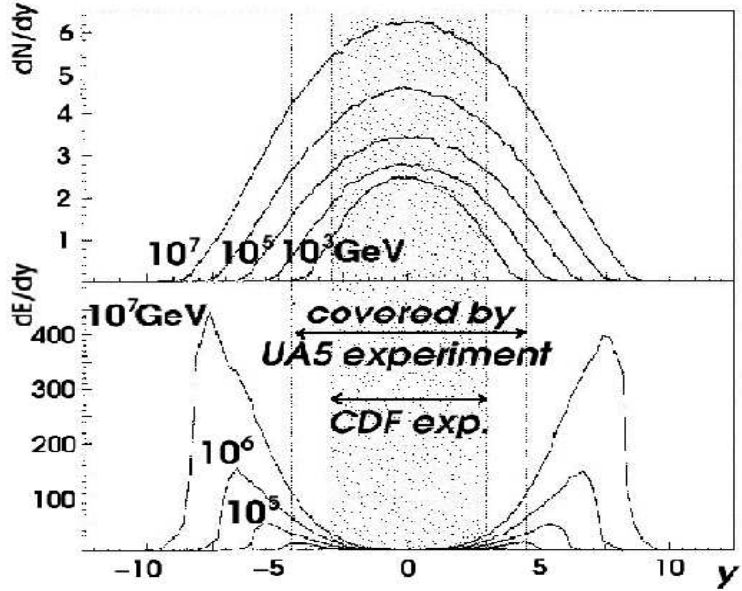,width=10.cm,height=8.cm}}
\caption{\em {Multiplicity distributions (up) and
energy distributions (down) vs. rapidity of secondary particles in high
energy $p-\bar{p}$ collisions for energy range 10$^3$ - 10$^7$ GeV in the rest
system of one of the primary particles. Vertical bands show the angular
acceptance of two collider experiments.}}
\end{center}
\label{rapidity}
\end{figure}

On the other hand, due to the lack of knowledge of the high energy
interactions from particle accelerators in this energy range, it is hard to
make comparisons between experimental results and simulations.
In fact, as has been pointed out by Jones~\cite{ref:jone} all these comparisons
require the knowledge of hadronic interactions at the highest energies and
over a wide angular range.
The highest energy reached currently is at Fermilab collider, 
studied with the CDF and D0 detectors, and 
corresponds to $\sim$ 2000 TeV proton collision with a stationary proton. The
LHC
collider currently under construction at CERN will extend this up to about
$\sim$10$^{17}$ eV.
(equivalent cosmic ray primary energy). These collider facilities will thus
cover most of the EAS energy region
of current interest in the contex of the "knee" and the primary composition
problems.
Unfortunately, this is not the case for the angular range (or rapidity)
coverage.
In the upper part of figure 3~\cite{ref:schatz},
the multiplicity distribution
of secondary particles for different primary energies is shown, together
with the rapidity range covered by the CDF experiment and UA5
experiments, which had the largest rapidity range achieved thus far among high
energy collider detectors.
It can be seen that most of the secondary particles are detected by both
detectors.
But in the lower part of figure 3, where the energy
distribution of secondary particles is displayed, it is clear that only
a small fraction of the primary energy is measured by these detectors, and
this fraction decreases with primary energy.
This effetc is even more important in the coordinate system where one (target)
nucleon is at
rest, i.e. the cosmic ray case.
Since the air shower development is dominated by the final state 
energy flow, i.e.
by this missing forward region, the most important information for EAS
simulation is not
provided by these collider experiments.

\section{Future prospects}

As is well known the energy region in the cosmic ray spectrum
at primary energies of $10^{15}-10^{16}$ eV presents the enigmatic
``knee''.
The origin of this change of slope of the primary cosmic rays spectrum is
still unresolved, even after a great deal of efforts has been dedicated with
increasingly
sophisticated equipments working in a variety of atmospheric depths in the
around the world.
The question about the origin of the ``knee'' is still open.
To know the reason of the existence of the knee is a major challenge
faced in the very near future. Is it due to a change in production and/or
acceleration mechanism at the source, propagation through interstellar space, or
perhaps
a change in the nature of the high energy interactions (as suggested by some
emulsion experiments at high altitude laboratories)?\\

The current thinking is that perhaps not just one reason but two (or more)
coincident
mechanisms are operating in this energy region to produce the "knee".
What should be the next steps for investigation of cosmic rays through this
energy range? One could expect that some answers to the problems
of nuclear interactions and multiple particle production
will be given by the current accelerators or
from the LHC experiments. However, as it is shown in figure 3,
we lose information in the very
forward directions with an increase in collider energy, unless different
detector architectures specifically sensitive
to small-angle (high rapidity) final states are utilized.\\
In order to fully exploit the great potential of high altitude
laboratories such as Chacaltaya,
the simultaneous measurement of different observables should be undertaken.
For example, consider an array built by 225 scintillation counters
arranged on a 7 m grid covering an area of $\sim 10^4~m^2$.
Since the atmospheric conditions at Chacaltaya for Cherenkov
measurements are optimal, a wide aperture Cherenkov array could complete
this EAS detector.
In the central part of the array an hadronic calorimeter should be installed.
This calorimeter should meet the following conditions:
a) enough thickness to realiably measure the energy of hadrons of up to 100 TeV,
b) large enough area to have
enough statistics, for primaries of at least $10^{16}$ eV, for 
example $100~m^2$, c) carpet detector on
the top (streamer tubes or RPC counters) to measure the fine structure of
the electromagnetic component of the associate shower, d) a tracking
system in the top layers to measure the arrival direction of surviving hadrons,
and
e) fine-grained tracking detectors in between layers, such as a silicon array,
scintillating fibers or emulsion chambers. The installation should also have
muon detectors
distributed around the EAS array and under the hadron calorimeter.\\
\subsection{Direct measurement of survival proton spectrum up to 100 TeV}
Up to the present the energy region up to some hundreds of TeV has been
investigated
by balloon- and satellite-borne instruments, and
above these energies only by ground-based air shower experiments which
are relatively insensitive to the primary nuclear composition.
The proposal below suggests how a direct measurement of the primary proton
spectrum could be achieved. \\
The surviving protons arriving at Chacaltaya suffer
an attenuation given by $N=N_o \cdot  e^{-540/ \lambda(E)}$
where $\lambda(E)$ is the nuclear interaction mean free path.
The energy of such events is measured by the calorimeter and they
are easily recognized by the lack of accompanying particles detected
either in the carpet or in the EAS array.
In addition, during moonless nights no accompanying Cherenkov light must be
detected with the Cherenkov array.
This confirms that no absorbed (lower energy) shower
in the upper atmosphere is associated with the surviving proton.
Even if this measurement is limited to lower proton energies and to a
small fraction of events, the information that we can obtain is of
crucial importance for background rejection.\\ 
Since only protons can reach the Chacaltaya level without interacting
in the atmosphere, the final result could be the measurement of the 
direct primary spectrum of protons provided, all the uncertainties, both in the
inelastic proton air nuclei cross section and diffraction scattering on air
nuclei be minimized.
With the proposed area of the calorimeter, there would be $\sim$ 80 events/year
for $E_p \geq 100 \ TeV$.
The measurement in this energy region is of crucial importance for the
calibration of the proton content in primary cosmic rays flux.\\
The number of expected events per
$100 \ m^2\cdot year \cdot sr$  in the energy range 1-100 TeV 
is shown in Figure 4.
The proposed direct measurement up to 100 TeV (or higher) is needed
to calibrate the indirect EAS measurement made
with the same detector at Chacaltaya.
In addition, these measurements of protons with good statistics can be
directly compared with balloon measurements.
\subsection{Pure events}
"Pure events" are those that interact about 2 collision m.f.p. above the
observation level and
have not suffered the further complicated process of cascade development.\\
A two nuclear interaction m.f.p. correspond to $\approx$ 4 radiation lengths
(r.l.), i.e. only $\approx$ 2 km above Chacaltaya.
Therefore, a very collimated hadronic jet is produced and the
electromagnetic cascade originated by $\pi^\circ$'s is at a very early
stage of development.
Therefore, this experiment provides a great opportunity to study
jet production in the very forward direction.
These events are impossible to study with collider detectors up to present.
\begin{figure}[htbp]
\begin{center}
\mbox{\epsfig{figure=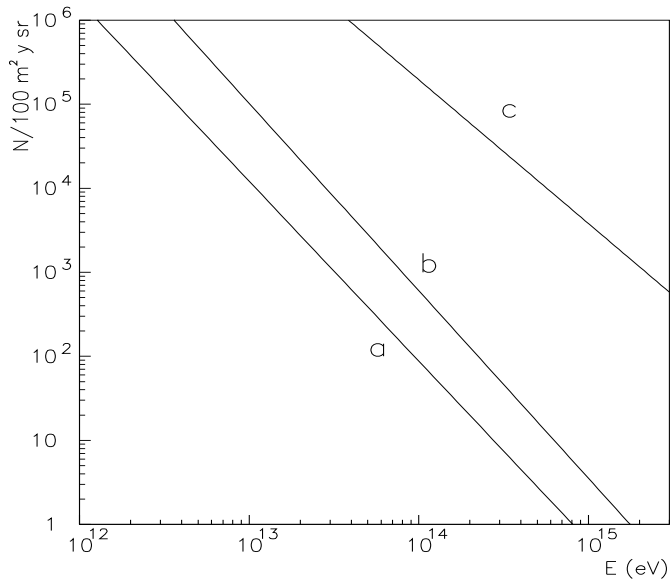,width=10.cm,height=8.cm}}
\caption{\em {The integral spectrum of survival primary protons (a), 
of pure events
(b) and of primary particles (c) at Chacaltaya per $100 \ m^2\cdot year$}}
\end{center}
\label{hecre}
\end{figure}

From figure 4, it is seen that few events/year are expected
at $10^{15}$ eV.
These events are also due to pure protons contained in the primary cosmic ray
spectrum.
In order to distinguish those events from the normal EAS
background we can use the information of the carpet detector as well as of the
EAS array and (at night) the Cherenkov light detectors.
In these events the electromagnetic cascade is at the very early
stage of its development and would have a very steep lateral distribution;
it should be well concentrated within the
$100~m^2$ carpet detector and not detected in the
($10^4~ m^2$) surrounding EAS array. From Figure 4 it is seen that about 100
events
per year are expected at over 200 TeV.\\
This would be a new approach to the study of the high-energy hadronic
interactions through jet production in the very forward direction together with
the associated electromagnetic component.

\subsection{Multihadron surviving events}
The detection of multihadrons ($\geq 2$) by the calorimeter could lead us
to study the primary composition in a semi-direct way.
In fact, Chacaltaya is at only $540 / \lambda(E)$  interaction m.f.p. and
14 r.l. from the top of atmosphere.
Therefore the study of events with a simultaneous measurements of the hadronic
energy and multiplicity, the associated EAS and the Cherenkov light
seems feasible at Chacaltaya.\\
Moreover, special types of events, like Centauro
and other exotic events like Chirons~\cite{ref:hase}
having a high hadronic multiplicity and high transverse momentum with no or low
electromagnetic component are still waiting to be understood.
The fine granularity detector (emulsion
chambers, silicon detector planes or scintillating fibers) inserted into 
the calorimeter
will measure such events. With silicon or scintillating fibers,
these events could be detected in real time, together with the associated 
EAS event.
Up to the present, such events have only been studied with emulsion 
technique.\\
A measurement of spectra of such multihadron events with energies up to
100 TeV would be
extremely helpful in order to better understand the phenomenology
of these higher-energy interactions.
A quantitative evaluation of such a study would be welcomed.

\subsection{Primary composition at the ``knee'' region}
As noted above, the ``knee'' in the energy
spectrum holds a key to the
understanding of the origin of cosmic ray. However, although a great
effort has been dedicated to
this problem over the past 40 years, confusion still reigns.\\
Lindner~\cite{ref:lind} proposed a
method based on the analysis of Cherenkov light and particle densities
registered relatively
close to the shower core; he shows that the $b$ vs. $X_{max}$ relationship,
(where $b$ is the ``slope'' of the lateral distribution of the Cherenkov light
and $X_{max} (gr/cm^2)$
is the depth of the maximum development of the Cherenkov shower),
does not depend on $A$, the primary mass, for a given model.\\
On the other hand, Procureur and Stamenov ~\cite{ref:proc} have introduced 
a new parameter $\alpha(r)$
such that a shower selected with fixed values of this parameter would be
generated by primaries with different masses but with the same primary energy.
It is worthwhile to note that $r$ for Chacaltaya is 35 m.\\
Both methods pursue an unbiased determination of the primary energy spectrum.
However, in order to significantly improve the unbiased determination of
the primary mass composition it has to be done by adding observables related
to the non-electromagnetic shower components.\\

Since at Chacaltaya elevation the showers are at the minimum of their
fluctuations for all components,
with the proposed combined techniques (EAS and Cherenkov array and
hadron-muon detector)
it could be possible to face the ``knee'' problem in a definitive way.
In particular, for lower energy showers the measurement
will overlap direct results obtained by satellites or balloons.
This possibility has a
basic importance because, up to now, EAS data have not been calibrated
against direct measurements.
\section{conclusions}

In this paper we have
shown that a possible experiment could be done at Chacaltaya that 
would have a
great potential, thanks to its high altitude location.
This presents a very great opportunity for cosmic ray physicists
to exploit the
unique conditions of this very high altitude mountain laboratory.\\
A suitable design of a collection of detectors,
perhaps modeled on that sketched here and undertaken by an international
collaboration, would demostrate the unique possibility to provide
very significant cosmic ray physics results.

\acknowledgments
Many people from several countries have contribuited to the revival of 
Chacaltaya Laboratory through their letters, conversations and discussions. 
O.S. would specialy like to thank the following: A. Watson, from Leeds 
University, E. Lorenz from HEGRA experiment,
G. Schatz and H. Rebel from KASCADE experiment, A. Ohsawa from ICRR,
University of Tokyo, for their 
continued interest in this work about Chacaltaya.

\end{document}